\documentclass[prd,showpacs]{revtex4}
\usepackage[utf8]{inputenc}
\usepackage[T1]{fontenc}
\usepackage{amsmath}
\usepackage{amssymb}
\usepackage{graphicx}
\usepackage{subfigure}
\usepackage{mathrsfs}
\usepackage{microtype}
\usepackage{xspace}
\usepackage{mathtools}

\allowdisplaybreaks
\def\jf#1{\relax}

\newcommand*{\scri}{\mathscr{I}}
\newcommand*{\dd}{\mathrm{d}}

\renewcommand*{\i}{\mathrm{i}}

\newcommand*{\del}{\partial}
\newcommand*{\RR}{\mathbb{R}}
\newcommand*{\eps}{\epsilon}

\newcommand*{\xib}{\bar\xi}
\newcommand*{\etab}{\bar\eta}

\newcommand*{\gammab}{\bar\gamma}
\newcommand*{\epsb}{\bar\eps}

\newcommand*{\sigmab}{\bar\sigma}

\newcommand*{\R}[1]{\mbox{\sffamily\ifcase#1\relax\or I\or II\or III\or IV\fi}\xspace}

\begin{document}

\title{Numerical evolution of plane gravitational waves in the Friedrich-Nagy gauge}
\author{J\"org Frauendiener}
\email{joergf@maths.otago.ac.nz}
\author{Chris Stevens}
\email{cstevens@maths.otago.ac.nz}
\author{Ben Whale}
\email{bwhale@maths.otago.ac.nz}
\affiliation{Department of Mathematics and Statistics, University of Otago, Dunedin 9010, New Zealand}
\date{\today}

\begin{abstract}
The first proof of well-posedness of an initial boundary value problem for the Einstein equations was given in 1999 by Friedrich and Nagy. They used a frame formalism with a particular gauge for formulating the equations. This `Friedrich-Nagy' (FN) gauge has never been implemented for use in numerical simulations before because it was deemed too complicated. In this paper we present an implementation of the FN gauge for systems with two commuting space-like Killing vectors. We investigate the numerical performance of this formulation for plane wave space-times, reproducing the well-known Khan-Penrose solution for colliding impulsive plane waves and exhibiting a gravitational wave `ping-pong'.
\end{abstract}

\maketitle

\section{Introduction}
\label{sec:introduction}

The initial boundary value problem (IBVP) for the Einstein equations is the basic tool for most numerical investigations of solutions to these equations. The main reason for this is the fact that infinite physical systems must be reduced to a finite size by the introduction of an artificial boundary (see~\cite{Frauendiener:2004te,Winicour:2001tn} for other situations). Then one is faced with the question as to which boundary conditions can be imposed so that the mathematical problem is well-posed. That is to say, for which class of boundary conditions does there exist a unique solution of the Einstein equations which assumes the given values on the initial data hyper-surface and the boundary? This ignores the problem of whether these conditions are also physical.

While there had been discussions of the solutions to Einstein's equations in the neighbourhood of a time-like hyper-surface before (see~\cite{Bartnik:1997jh,Kijowski:1997wt,Tamburino:1966vd}), the first paper to discuss the IBVP for the Einstein equations from the point of view of well-posedness of a system of PDEs is by H.~Friedrich and G.~Nagy~\cite{Friedrich:1999dc}. In~\cite{Friedrich:1995uf}, Friedrich discussed the asymptotic IBVP for the Einstein equations with negative cosmological constant, where null-infinity~$\scri$ is time-like and can be used to impose boundary conditions for the asymptotic fall-off of the gravitational field. However, this situation is special in the sense that null-infinity is a geometrically distinguished hyper-surface and consequently the data are of a very particular type as required by the geometry of $\scri$.

The discussion in~\cite{Friedrich:1999dc} is based on a formulation of the Einstein equations as a system of $1^\text{st}$-order PDEs for geometric quantities including the Weyl tensor and the authors present a class of boundary conditions for the components of the Weyl tensor in a particular adapted gauge (the FN gauge), for which the IBVP is well-posed.

However, the standard numerical codes usually employ second-order formulations and as a consequence the FN gauge and boundary conditions have never been implemented numerically. Meanwhile, there are other treatments of the IBVP (see the review paper~\cite{Sarbach:2011tq} and references therein) based on the generalised harmonic gauge, which puts the Einstein equations into a system of ($2^\text{nd}$-order) wave equations.

In this work, we want to study the numerical performance of the FN gauge and the related boundary conditions in a very simple example. To this end we simplify the full Einstein equations by imposing symmetries. Since the assumption of spherical symmetry precludes the existence of gravitational waves we consider solutions which possess two commuting space-like Killing vectors. These generate the action of a 2-dimensional abelian group of isometries and we assume that the orbits are diffeomorphic to $\RR^2$. Under these conditions our system describes plane waves. Such systems have been studied thoroughly in the literature over the years, mostly in the context of exact solutions and their generation techniques (see the book by Griffiths~\cite{Griffiths:1991vx} and references therein). Of particular interest has been the case of colliding plane waves.

In 1971, K.~Khan and R.~Penrose~\cite{Khan:1971ew} published the first exact solution describing two impulsive plane gravitational waves, which approach each other, collide head-on---thereby interacting non-linearly---and then separate again. This prototype solution has been generalised later to allow for the waves to have different polarisations and also include matter waves. 

As the first application of our numerical code we show how to numerically approach the Khan-Penrose solution by approximating the delta-function profile of the ingoing waves with a sequence of smooth bump-functions. Another application is the gravitational wave `ping-pong', where an incoming gravitational wave is reflected back and forth at the boundaries using boundary conditions from the admissible class exhibited by Friedrich and Nagy. While these boundary conditions lead to a mathematically well-posed IBVP, they have no physical meaning, because there is no known mechanism which would reflect gravitational waves.

The structure of the paper is as follows. In sec.~\ref{sec:equations} we present the equations incorporating the symmetry and the FN gauge. Their numerical implementation and basic tests are discussed in sec.~\ref{sec:numer-setup}. Sec.~\ref{sec:khan-penr-solut} is devoted to the reproduction of the Khan-Penrose solution and in sec.~\ref{sec:grav-wave-ping} we present a brief description of the gravitational wave ping-pong. A short summary ends the paper.

\section{Derivation of the field equations}
\label{sec:equations}

\paragraph{Symmetries and gauge conditions}
In deriving the equations we largely follow the derivation given in~\cite{Griffiths:1991vx}. We assume that the space-time $(M, g)$ admits the action of the 2-dimensional translation group by isometries. We assume that the orbits are topologically $\RR^2$ and that they foliate~$M$. They are intrinsically flat, and so we refer to them as `planes', even though they do not necessarily admit the action of the full 2-dimensional Euclidean group. In fact, if they did, there would be no gravitational waves present in the space-time. We can introduce local coordinates $(x,y)$ in the planes such that the Killing vectors are represented as the coordinate derivatives $\del_x$ and $\del_y$.

Next, we introduce local coordinates $t$ and $z$, which are constant on the planes, and locally characterise them uniquely; we write $\Pi_{tz}$ for the plane with constant values of $t$ and $z$. We fix these coordinates further by imposing gauge conditions. Here, we impose the Friedrich-Nagy gauge conditions, see~\cite{Friedrich:1999dc}. First, we fix a space-like hyper-surface $S_0$ to specify initial data for the IBVP. We assume that the Killing vectors are tangent to $S_0$, such that $S_0$ is foliated by planes, i.e. it is homeomorphic to $I\times\RR^2$ for some finite interval $I$. Let $z_l<z_r \in I$, and consider time-like hyper-surfaces $T_l$ and $T_r$, also ruled by planes, which intersect $S_0$ in two planes $\Pi_{0l}:=S_0 \cap T_l$ and $\Pi_{0r}:=S_0 \cap T_r$. 

Next, we choose a time-like unit vector-field $t^a$, invariant under the group action, so that it is tangent to $T_l$ and $T_r$ and otherwise arbitrary in $M$. We fix the time coordinate $t$ by requiring that it be the parameter along this vector-field. Thus, $t$ is the proper time for the observers with 4-velocity $t^a$. Due to the symmetry, the $x$ and $y$ coordinates are constant along $t^a$ and we fix the $z$-coordinate also by this requirement, such that $t^a\nabla_a z = 0$.

We also fix a global tetrad on $M$ as follows. Let $z^a$ be the (space-like) unit normal to the hyper-surfaces $T_z$ of constant $z$. Note that this vector will be the outward normal at one end of the interval and the inward normal at the other end. We choose it so that it is outward pointing at the right end of the interval. Furthermore, we choose two mutually orthogonal space-like unit vectors $x^a$ and $y^a$ tangent to the planes. Together with $t^a$, these form a tetrad field $(t^a, x^a, y^a, z^a)$ on (a neighbourhood of)~$M$. Note that the tetrad vectors $x^a$ and $y^a$ are unique up to a rotation within the planes. In~\cite{Friedrich:1999dc}, they were fixed by the requirement that they be Fermi transported within the hyper-surfaces $T_z$. Here, we allow for the arbitrary rotation in the form of a gauge source function, see below.

The position of the hyper-surfaces $T_z$ within $M$ will be determined in terms of their mean curvature $\chi$, which is specified as an arbitrary function of the coordinates $(t,z)$, i.e. an invariant gauge source function.

\paragraph{Newman-Penrose equations}

In this article we are interested in solutions of Einstein's equations in vacuum,
\[
R_{ab} = 0.
\]
We use the Newman-Penrose formalism~\cite{Newman:1962ue} to derive the complete set of field equations. First, we choose an adapted null-tetrad as follows. At each point of the plane $\Pi_{tz}$ there exist two unique null-directions given by future-pointing null-vectors $l^a$ and $n^a$, orthogonal to the plane and assumed to be normalised against each other $l_an^a = 1$\footnote{We will use the notation and conventions of Penrose and Rindler\cite{Penrose:1984wm} throughout.}. To eliminate the remaining boost freedom we require that $t^a = \frac1{\sqrt2} (l^a + n^a)$. In addition to the two real null-vectors, we choose the complex null-vector $m^a = \frac1{\sqrt2}\left( x^a + \i y^a\right)$. 

Our conditions imply
\[
t^a\nabla_a = \del_t, \quad \text{ and } z_a = -\frac1A \nabla_a z \implies z^a\nabla_a  = B \del_t + A \del_z + X \del_x + Y \del_y
\]
for invariant functions $A(t,z)$, $B(t,z)$, $X(t,z)$ and $Y(t,z)$. Using the freedom in the choice of coordinates $x$ and $y$ within each plane, $x \mapsto a x + b y$, $y \mapsto c x + d y$ with invariant functions $a(t,z)$, $b(t,z)$, $c(t,z)$, and $d(t,z)$, we can eliminate the functions $X$ and $Y$.  Therefore, the  null-tetrad $(l^a, n^a,m^a)$ can be represented in terms of the directional derivatives
\begin{align}
  D &= l^a \nabla_a = \frac1{\sqrt2} \left((1+B) \del_t + A
    \del_z \right),\label{eq:1} \\ 
  D'&= n^a \nabla_a = \frac1{\sqrt2} \left((1-B)
    \del_t - A \del_z\right),\label{eq:2} \\
  \delta &= m^a\nabla_a = \xi \del_x + \eta    \del_y, \label{eq:3}
\end{align}
in terms of the coordinates $(t,x,y,z)$. The complex space-like vector $m^a$ is defined in terms of invariant complex-valued functions $\xi(t,z)$ and $\eta(t,z)$, but it is only defined up to a $U(1)$-valued function of all the coordinates.

In these coordinates, the metric assumes the form
\begin{equation}
  \label{eq:4}
  g = \left(\dd t - \frac{B}{A}\, \dd z\right)^2 - \frac{1}{A^2}\, \dd z^2 + \frac2{(\xi\etab - \xib\eta)^2} \left(\eta\, \dd x - \xi \dd y \right) \left(\etab\, \dd x - \xib \dd y \right).
\end{equation}
We now use the commutator relations between the directional derivatives (see~\cite{Penrose:1984wm}, eqn. (4.11.11)) applied to the coordinates to get the following relationships between the tetrad functions and the spin-coefficients
\[
\begin{gathered}
  \rho=\bar\rho,\quad
  \rho'=\bar\rho',\quad
  \kappa'= 0,\quad
  \kappa = 0,\quad 
  \alpha = 0, \quad
  \beta=0,\quad \tau=0,\quad \tau'=0,\\
  (D+D')B = (\gamma + \gammab + \eps + \epsb) + (\gamma + \gammab -
  \eps - \epsb) B,\\
  (D+D')A = (\gamma + \gammab - \eps - \epsb)A,\\ 
  (D+D')X = (\gamma + \gammab - \eps - \epsb)X,\\ 
  (D+D')Y = (\gamma + \gammab - \eps - \epsb)Y,\\ 
  D\xi = \sigma\xib + (\eps-\epsb+\rho)\xi,\qquad
  D'\xi = \sigmab'\xib + (\gamma-\gammab+\rho')\xi,\\
  D\eta = \sigma\etab + (\eps-\epsb+\rho)\eta,\qquad
  D'\eta = \sigmab'\etab + (\gamma-\gammab+\rho')\eta.
\end{gathered}
\]
The next set of equations comes from the curvature equations (see~\cite{Penrose:1984wm}, eqn. (4.11.12)), which under the given simplifications and with the vacuum equations, $\Phi_{ik} = 0$ and $\Lambda = 0$, read
\[
\begin{gathered}
  D\rho = \rho^2 + \sigma \sigmab + \rho(\eps+\epsb),\\
  D'\rho = 2 \rho\rho'  + \rho(\gamma+\gammab),\\[1ex]
  D\rho' = 2 \rho\rho'  - \rho'(\eps+\epsb),\\
  D'\rho' = \rho'^2 + \sigma' \sigmab' - \rho'(\gamma+\gammab),\\[1ex]
  D\sigma = 2\rho\sigma + \sigma(3\eps-\epsb) + \Psi_0,\\
  D'\sigma = \rho'\sigma + \rho\sigmab' + \sigma(3\gamma-\gammab),\\[1ex]
  D\sigma' = \rho\sigma' + \rho'\sigmab - \sigma'(3\eps-\epsb),\\
  D'\sigma' = 2\rho'\sigma' - \sigma'(3\gamma-\gammab) +
  \Psi_4,\\[1ex]
  D\gamma - D'\eps = -\rho\rho' + \sigma\sigma' - \eps(\gamma+\gammab)
  - \gamma(\eps+\epsb) 
\end{gathered}
\]
together with algebraic conditions for the Weyl tensor components $\Psi_1$, $\Psi_2$ and~$\Psi_3$
\begin{equation}
\Psi_1 = 0, \qquad \Psi_2 = \sigma\sigma' - \rho \rho', \qquad \Psi_3 = 0.\label{eq:23}
\end{equation}
The second of these is simply a consequence of the vanishing Gauß curvature of the planes and the vanishing of the Ricci tensor.

The final set of equations comes from the Bianchi identities, regarded as equations for the Weyl tensor components
\[
\begin{aligned}
  D'\Psi_0 &= 3 \sigma \Psi_2 + (\rho'+4\gamma)\Psi_0,\\
  D\Psi_4 &= 3 \sigma' \Psi_2 + (\rho-4\eps)\Psi_4.
\end{aligned}
\]
We note that the equations for the functions $\xi$ and $\eta$ decouple from the remaining system in the sense that they can be integrated separately once the remaining equations have been solved. Therefore, we will ignore them for the time being. 

Furthermore, we note that we do not have enough equations to determine the spin-coefficients $\eps$ and $\gamma$. In order to obtain the missing information, we turn to the gauge conditions that we imposed. First, we determine the mean curvature of the hyper-surfaces $T_z$ in terms of the spin-coefficients:
\begin{equation}
\chi = (g^{ab} + z^a z^b)\nabla_a z_b = \frac1{\sqrt2}\left( \eps + \epsb + \gamma + \gammab - 2 \rho + 2 \rho'\right).\label{eq:5}
\end{equation}
It is easily seen that Fermi transport $\mathbf{F}$ of $x^a$ and $y^a$ (and, hence, of $m^a$) along $t^a$ within $T_z$ reduces to parallel transport due to the symmetry assumptions. Therefore, we can write
\begin{equation}
\mathbf{F} m^a = t^c\nabla_c m^a = \frac1{\sqrt2} \left( \eps - \epsb + \gamma - \gammab\right)m^a.\label{eq:6}
\end{equation}
The remaining gauge freedom in the choice of $m^a$ can be expressed by the equation
\[
\mathbf{F}m^a = \i f m^a
\]
for an arbitrary real-valued invariant function $f(t,z)$. Combining~\eqref{eq:5} and~\eqref{eq:6}, we obtain
\[
\eps + \gamma =  \rho - \rho' + F,
\]
where $F(t,z)$ is an invariant complex-valued function, with $\sqrt2 F = \chi + \i f$. Introducing a new complex-valued function $\mu(t,z)$ by defining
\[
\mu = \gamma - \eps,
\]
we can express $\eps$ and $\gamma$ in terms of $\mu$ and $F$ as follows
\[
\eps = \frac12 (\rho - \rho' + F - \mu), \qquad
\gamma = \frac12 (\rho - \rho' + F + \mu).
\]
Inserting these expressions into the last of the curvature equations, we obtain an equation for the function $\mu$
\begin{multline}
  (D+D') \mu = \mu^2 - 3 (\rho - \rho')^2 - (\rho - \rho')(\bar F + 3
  F) + (\mu + \bar \mu) (\rho + \rho') + \mu\bar \mu\\ - \sigma
  \bar\sigma - \sigma'\bar\sigma' + 2 \sigma\sigma' - F^2 - F \bar F -
  DF + D'F\label{eq:7}.
\end{multline}

\paragraph{Evolution and constraint equations}

The final step is the splitting of the equations obtained in the previous paragraph into evolution equations and constraints. Combining the equations appropriately, we obtain the evolution equations:
\begin{subequations}
\label{eq:evolution}
  \begin{align}
    \label{eq:8}
    \sqrt2 \del_t A &= (\mu + \bar\mu)\,A,\\
    \sqrt2 \del_t B &= (2\rho - 2\rho' + F + \bar F) + (\mu + \bar\mu) B,\label{eq:9}\\
    \sqrt2 \del_t \rho &= 3\rho^2 + \sigma \sigmab + \rho(F + \bar F),\label{eq:10}\\
    \sqrt2 \del_t \rho' &= 3\rho^{\prime2}  + \sigma' \sigmab' - \rho'(F + \bar F),\label{eq:11}\\
    \sqrt2 \del_t \sigma &= 4\rho\sigma - \rho'\sigma + \rho\sigmab' + \sigma(3F - \bar F) + \Psi_0,\label{eq:12}\\
    \sqrt2 \del_t \sigma' &= 4\rho'\sigma' - \rho\sigma' +
    \rho'\sigmab - \sigma'(3 F - \bar F) +
    \Psi_4,\label{eq:13}\\
    \sqrt2 \del_t \mu &= \mu^2 + \mu\bar \mu - 3 (\rho - \rho')^2 +
    (\mu + \bar \mu) (\rho + \rho') - \sigma \bar\sigma -
    \sigma'\bar\sigma' + 2 \sigma\sigma'\nonumber\\ & - (\rho -
    \rho')(\bar F + 3
    F) - F^2 - F \bar F - \sqrt2 A\del_z F - \sqrt2 B\del_t F,\label{eq:14}\\
    (1&-B) \del_t \Psi_0 - A \del_z \Psi_0 = \sqrt2 \left(3 \sigma \Psi_2 + (2\rho - \rho' + 2 F + 2\mu)\Psi_0\right),\label{eq:15}\\
    (1&+B) \del_t \Psi_4 + A \del_z \Psi_4= \sqrt2 \left(3 \sigma'
      \Psi_2 + (2\rho' - \rho - 2 F - 2\mu)\Psi_4\right),\label{eq:16}
  \end{align}
\end{subequations}
and the remaining equations can be written as the vanishing of constraint quantities:
\begin{subequations}
  \label{eq:constraints}
  \begin{align}
    \label{eq:17}
    0=C_1 &:= \sqrt2 A\del_z\rho - (1 - 3 B) \rho^2 + 2\rho\rho' - (1 - B) \sigma\sigmab \nonumber \\
    &\hskip12em+ \rho (\mu + \bar\mu) + \rho B (F+\bar F),\\
    0=C_2 &:= \sqrt2 A\del_z\rho' + (1 + 3 B) {\rho'}^2 - 2\rho\rho' + (1 + B) \sigma'\sigmab' \nonumber\\
    &\hskip12em - \rho' ( \mu + \bar\mu ) - \rho' B (F+\bar F) ,\label{eq:18}\\
    0=C_3&:=\sqrt2 A\del_z\sigma + (1+B) \rho\sigmab' - 2 (1-2B)\rho\sigma + (1-B) \rho'\sigma  \nonumber\\ &\hskip8em + \sigma(3\mu - \bar\mu) + B\sigma(3F - \bar F) - (1-B) \Psi_0 ,\label{eq:19}\\
    0=C_4&:= \sqrt2 A\del_z\sigma' - (1-B) \rho'\sigmab + 2
    (1+2B)\rho'\sigma' - (1+B) \rho\sigma' \nonumber\\ &\hskip8em -
    \sigma'(3\mu - \bar\mu) - B\sigma'(3F - \bar F) + (1+B)
    \Psi_4. \label{eq:20}
  \end{align}
\end{subequations}
The evolution equations \eqref{eq:evolution} are a system for the tetrad components $A$ and $B$, the divergences $\rho$ and $\rho'$, the shears $\sigma$ and $\sigma'$, the auxiliary `spin-coefficient' $\mu$, and the two Weyl tensor components $\Psi_0$ and $\Psi_4$. All of the equations are advection equations, the first seven along the vector field $t^a$ and the last two along the null-vectors $n^a$ or $l^a$. The system is therefore symmetric hyperbolic and we obtain solutions for arbitrary initial conditions.

The second system \eqref{eq:constraints} can be read as four equations for the divergences and shears to be satisfied on every hyper-surface of constant $t$, the other functions ($A$, $B$, $\mu$, $F$, $\Psi_0$ and $\Psi_4$) being specified freely. Using the evolution equations, one can show that the constraint quantities $(C_1,\ldots,C_4)$ satisfy a linear system of ODEs of the form
\[
\frac{\dd}{\dd t} \mathbf{C} = M \mathbf{C} + N \mathbf{\bar C},
\]
where $M$ and $N$ are $4\times4$ complex matrices depending only on the unknowns. Thus, the constraints propagate. What is more important is the fact that the constraint quantities propagate within the hyper-surfaces $T_z$, in particular they do not cross boundaries given by constant values of $z$. This implies that we do not need to impose any special kind of `constraint preserving' boundary conditions. This is a special case of the theorem proven in~\cite{Friedrich:1999dc}. The only boundary conditions that are necessary for this system are boundary values for the Weyl tensor components $\Psi_0$ and $\Psi_4$. If $A>0$ and $|B|<1$, then $\Psi_0$ travels towards the left and needs a boundary condition at the right-hand side of the interval, while $\Psi_4$ travels to the right and needs a boundary condition on the left.

Taking these facts together we find that the two systems~\eqref{eq:evolution} and~\eqref{eq:constraints} combined admit a well-posed initial boundary value problem.

\section{Numerical setup}
\label{sec:numer-setup}

In our numerical code, we choose the interval $I=[-1,1]$ as our computational domain. Since most of the evolution equations are ODEs in time, only the equations for the Weyl tensor components need spatial discretisation. We approximate the spatial derivative $\del_z$ by a finite difference operator which has the summation by parts (SBP) property (see~\cite{Lehner:2004ha,Strand:1994ef}) at both ends of the interval. The time evolution is done with the standard 4th order Runge-Kutta method.

\paragraph{Boundary conditions}
\label{sec:boundary-conditions}

As discussed above the only boundary conditions we need are for $\Psi_0$ at $z=1$ and $\Psi_4$ at $z=-1$. We impose these conditions using the SAT method, which has been discussed in detail in~\cite{Schnetter:2006ku,Carpenter:1993uu,Carpenter:1993wh}. Our main application will be the case where one or two plane waves enter an initially flat domain. We achieve this by initially giving constant values to $A$ and $B$ and putting all other variables to zero. Under these conditions the constraint equations are satisfied initially.

\paragraph{Characteristic coordinates}
\label{sec:char-coord}

We will describe the evolution in terms of characteristic coordinates $(u,v)$, which we define by the equations
\[
D'u = 0, \qquad Dv = 0.
\]
We add these advection equations to the evolution system, so that we compute these coordinates together with the geometry. The boundary conditions for $u$ and $v$ are
\[
u(t,-1) = v(t, 1) = \frac1{\sqrt2} t
\]
at the boundaries $z=\pm1$. These have the consequence that $u(0,-1) = v(0,1) = 0$. With appropriate initial conditions, discussed later, the characteristic coordinates are fixed uniquely, since they are determined in terms of the conformal structure of the 2-dimensional space of orbits. We use these coordinates  (or their time-like and space-like counterparts $T=T_0 + (v+u)/\sqrt2$ and $Z=(v-u)/\sqrt2$) to describe the solutions in an (almost) invariant way.

\paragraph{Code tests}
\label{sec:code-tests}

\begin{figure}[htb]
  \centering
  \includegraphics[width=0.8\textwidth]{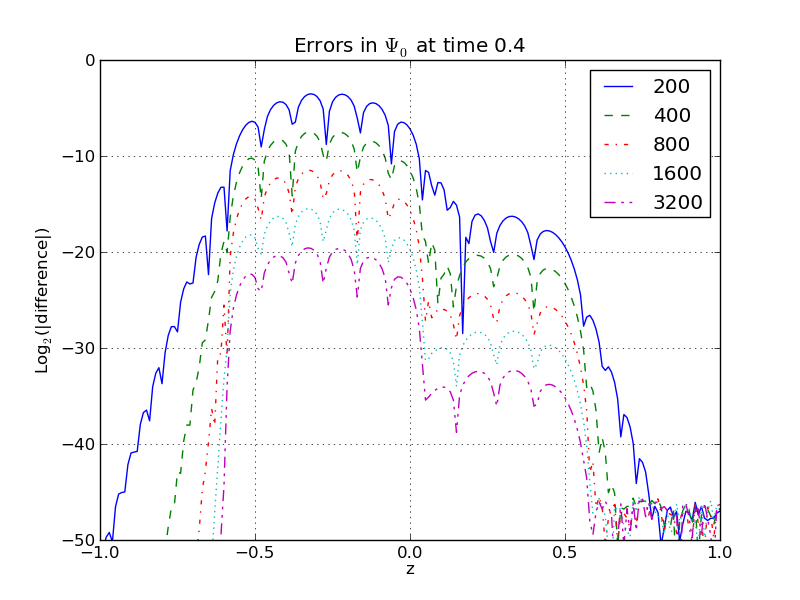}  
  \caption{Convergence plot of $\Psi_0$ at time 0.4}
  \label{fig:convergence}
\end{figure}

In order to show the validity of our code, we tested it for convergence. We ran our first scenario, described in the following section, with weak waves coming in from both boundaries for almost one crossing time, using an increasing number of grid points $N \in \{200, 400, 800, 1600, 3200, 6400\}$, and computed the difference to an evolution using the maximal number of $6400$ grid points. The results are shown in Fig.~\ref{fig:convergence},
 where we can see the expected $4^{\text{th}}$ order convergence.

\begin{figure}[htb]
  \centering
  \includegraphics[width=0.8\textwidth]{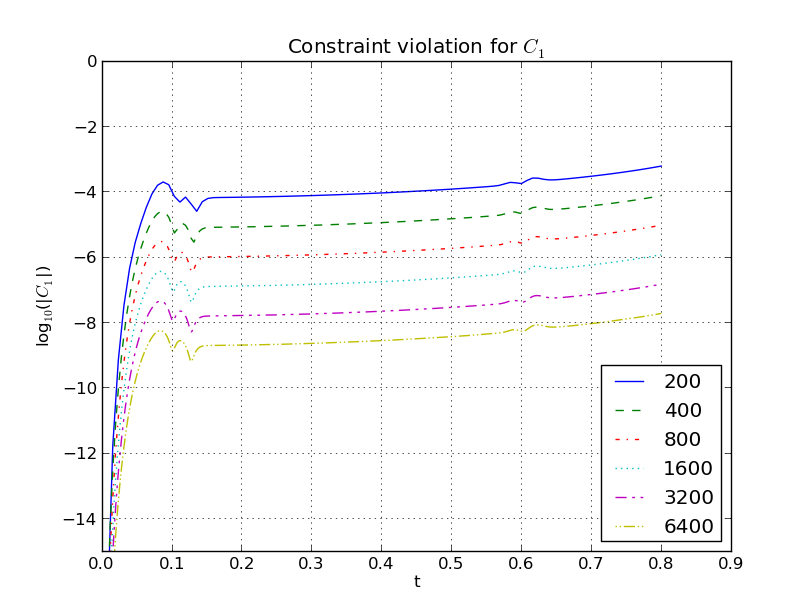}
  \caption{$Log_{10}$ of the L2-norm of $|C_1|$ over the course of the simulation}
  \label{fig:constraints}
\end{figure}

With the same setup, we also checked for constraint violations. In Fig.~\ref{fig:constraints}
we display the evolution of the constraint quantities over time. Again we see the expected behaviour. We should point out here that we have not (yet) explored the possibility of modifying the evolution equations by adding combinations of the constraint quantities in order to obtain an evolution system which damps the constraints on the fly.

\paragraph{Exploring the gauge source function}
\label{sec:expl-gauge-source}

To get some feeling for the effect of the gauge source function $F$, we consider the simple case of Minkowski space with metric $g=\dd T^2 - \dd Z^2 - \dd X^2 - \dd Y^2$. We take the coordinate vectors $\del_X$ and $\del_Y$ as the Killing vectors so that the planes are given by constant values of the coordinates $T$ and $Z$, as before. We will not concern ourselves with the imaginary part of $F$, since this determines the orientation of the frame vectors spanning the planes and therefore only affects the phases of the complex spin-weighted quantities, such as the shears and the curvature scalars.

The real part of $F$ determines the mean curvature $\chi$ of the constant $z$ hyper-surfaces. Projecting these onto the $(T,Z)$ plane we obtain a curve. We write the curve in parametrised form as $(T(t),Z(t))$ and we assume that $T(0)=0$ and $Z(0)=z$. Taking the coordinate $t$ along the curve as proper time (with $\dot T>0$), we have
\[
\del_t = \gamma(t) \del_T + v(t) \gamma(t) \del_Z, \quad \text{where } \gamma = \dot T,\quad v = \frac{\dot Z}{\dot T},  \quad\text{and } \gamma^2(1-v^2)=1.
\]
The normal vector $\mathbf{n}$ to the curve pointing towards increasing $Z$ is
\[
\mathbf{n}(t) = \gamma(t) \left(v(t)\del_T+\del_Z\right).
\]
The induced metric on this curve is $\dd t^2$, and it is easy to compute the extrinsic curvature, which comes out to be
\[
K = \frac12 \mathscr{L}_{\mathbf{n}} g = \frac{\dot v}{1-v^2}\dd t^2.
\]
Then, the mean curvature is 
\[
\chi(t) = \frac{\dot v}{1-v^2}.
\]
Integrating these relationships for constant $\chi>0$ and subject to the additional condition  $\dot Z(0)=0$, we obtain the hyperbolic motion
\[
T(t) = \frac1\chi \left( \cosh(\chi t) - 1\right), \qquad 
Z(t) = z + \frac1\chi \sinh(\chi t).
\]
Thus, in this simplified case, the mean curvature plays the role of the acceleration of the lines of constant $z$. In more complicated cases, even within the plane symmetric space-times, the relationship is not so clear, and the steering of the curves becomes much more indirect. Without the symmetry, prescribing the mean curvature of a constant $z$ hyper-surface influences the entire hyper-surface, and it is difficult to predetermine how the hyper-surface will behave for different values of the mean curvature, apart from the more or less obvious fact that it will bend outward for one choice of its sign and inward for the other choice.

\section{The Khan-Penrose solution}
\label{sec:khan-penr-solut}

\begin{figure}[htb]
  \centering
  \includegraphics[width=0.5\linewidth]{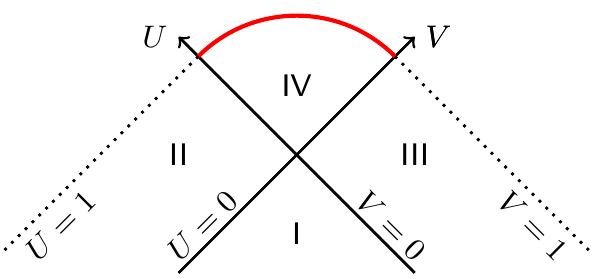}
  \caption{The global structure of the Khan-Penrose solution as projected onto the $(U,V)$-plane. Each point corresponds to one plane of symmetry. The dotted lines mark the coordinate fold singularity, while the thick line gives the location of the curvature singularity.}
  \label{fig:singstruct}
\end{figure}

In 1971 Khan and Penrose~\cite{Khan:1971ew} published an exact solution of Einstein's vacuum equations describing the head-on collision of two impulsive gravitational waves, which approach each other through a flat region of Minkowski space, interact, and then separate again. The metric that describes this process is
\begin{equation}
  \label{eq:21}
  \begin{multlined}
    g = \frac{2T^3 \, \dd U \dd V}{RW(PQ + RW)^2} - T^2 \left(\frac{R
        + Q}{R - Q}\right) \left(\frac{W + P}{W - P}\right)\,\dd X^2 \\
   - T^2 \left(\frac{R - Q}{R + Q}\right) \left(\frac{W - P}{W +
        P}\right)\,\dd Y^2.
  \end{multlined}
\end{equation}
The functions $P$, $Q$, $R$, $W$ and $T$ are defined as follows
\[
\begin{gathered}
  P = U\theta(U),\quad Q = V\theta(V),\quad R = \sqrt{1-P^2}, \quad W
  = \sqrt{1-Q^2}, \\ T = \sqrt{1 - P^2 - Q^2},
\end{gathered}
\]
with $\theta$ being the Heaviside step function, 
\[
\theta(x) =
\begin{cases}
  0 & x \le 0, \\
  1 & x > 0.
\end{cases}
\]
The form of this metric suggests that the space-time splits in a natural way into four regions, traditionally labelled \R1, \R2, \R3, and \R4 defined by the signs of $U$ and $V$
\[
\begin{cases}
  U\le0,\quad V\le0 & \text{in region \R1},\\
  U>0,\quad V\le0 & \text{in region \R2},\\
  U\le0,\quad V>0 & \text{in region \R3},\\
  U>0,\quad V>0 & \text{in region \R4}.
\end{cases}
\]
In region \R1, we have $P=Q=0$ and $R=W=T=1$, so that the metric becomes
\[
    g_{\R1} = 2\, \dd U \dd V - \dd X^2 - \dd Y^2,
\]
the flat metric in double null coordinates. In region \R2, the metric is
\[
    g_{\R2} = 2\, \dd U \dd V - (1 + U)^2\,\dd X^2 - (1 - U)^2\,\dd Y^2,
\]
and similarly in region \R3 (with $U$ and $V$ interchanged). Both these regions are also flat, but they contain a coordinate singularity (fold singularity) at $U=1$ resp. $V=1$. Region \R4 is the interaction region, which is not flat. In fact, Khan and Penrose show that region \R4 contains a space-like curvature singularity, which is located along the line $U^2 + V^2 = 1$, see~Fig.~\ref{fig:singstruct}. The structure of the singularities in the Khan-Penrose solution and its causal properties are discussed in great detail in~\cite{Matzner:1984bx}, see also~\cite{Griffiths:1991vx}.

The two waves come in along the lines $U=0$ and $V=0$, colliding in the central point. They travel through each other, one wave lensing the other, and separate again. After the collision, the waves maintain a strengthened impulsive component in their profile but also pick up a tail. The exact expressions for the Weyl spinor components and the curvature invariant $I$ in region \R4 are quite complicated. We give them in Appendix~\ref{sec:exact-kp} .

In order to reproduce the Khan-Penrose solution, we approximate the impulsive ingoing profile by a sequence of profiles which approximate the $\delta$-functions in the exact solutions. We have chosen appropriate members from the family of bump-functions
\[
\rho_l(x) =
\begin{cases}
  \frac{128}{35 l}\sin^8(\pi x/l) &  0 \le x \le l\\
  0 & \text{otherwise}
\end{cases},
\]
which satisfy the condition $\int_{-\infty}^\infty \rho_l = 1$ and have shrinking support, approaching $\{0\}$ for $l\to 0$. 

Our numerical setup is as described in sec.~\ref{sec:numer-setup}. The computational domain is the interval $[-1,1]$. We specify initial data for $t=0$, which correspond to the initial region (\R1) being flat, i.e. $B=0$, $\xi=\i\eta=1/\sqrt2$, and $A$ is constant. Choosing $A=1/L$ yields an initial length $2L$ of the domain. All other functions in the system vanish for $t=0$. We also specify the initial values for the characteristic coordinates $v = (-1+z)/\sqrt2$ and $u = (-1-z)/\sqrt2$.

On the boundaries, we specify values for $u$ and $v$ as given in~sec.~\ref{sec:numer-setup}, as well as the profiles of the ingoing waves. Thus, on the left boundary we put for some value of $l$
\[
u(t,-1) = \frac{t}{\sqrt2}, \qquad \Psi_4(t,-1) = \rho_l(u(t,-1)),
\]
and on the right boundary
\[
v(t,1) = \frac{t}{\sqrt2}, \qquad \Psi_0(t,1) = \rho_l(v(t,1)).
\]
To fix the coordinates completely, we need to fix a gauge source function $F(t,z)$. However, since we will analyse the geometric quantities in their dependence on the characteristic coordinates, which are computed simultaneously, the influence of the gauge is quite small. It affects the size and shape of the patch of space-time which is covered by the coordinates, but not the dependence of the geometric quantities on the characteristic coordinates. We have run the code with both $F=0$ and with $F$ computed from the explicit representation of the Khan-Penrose metric~\eqref{eq:21}, and did not find other differences. So here, we present our results obtained with $F=0$.

\begin{figure}[htb]
  \centering
  \includegraphics[width=0.8\textwidth]{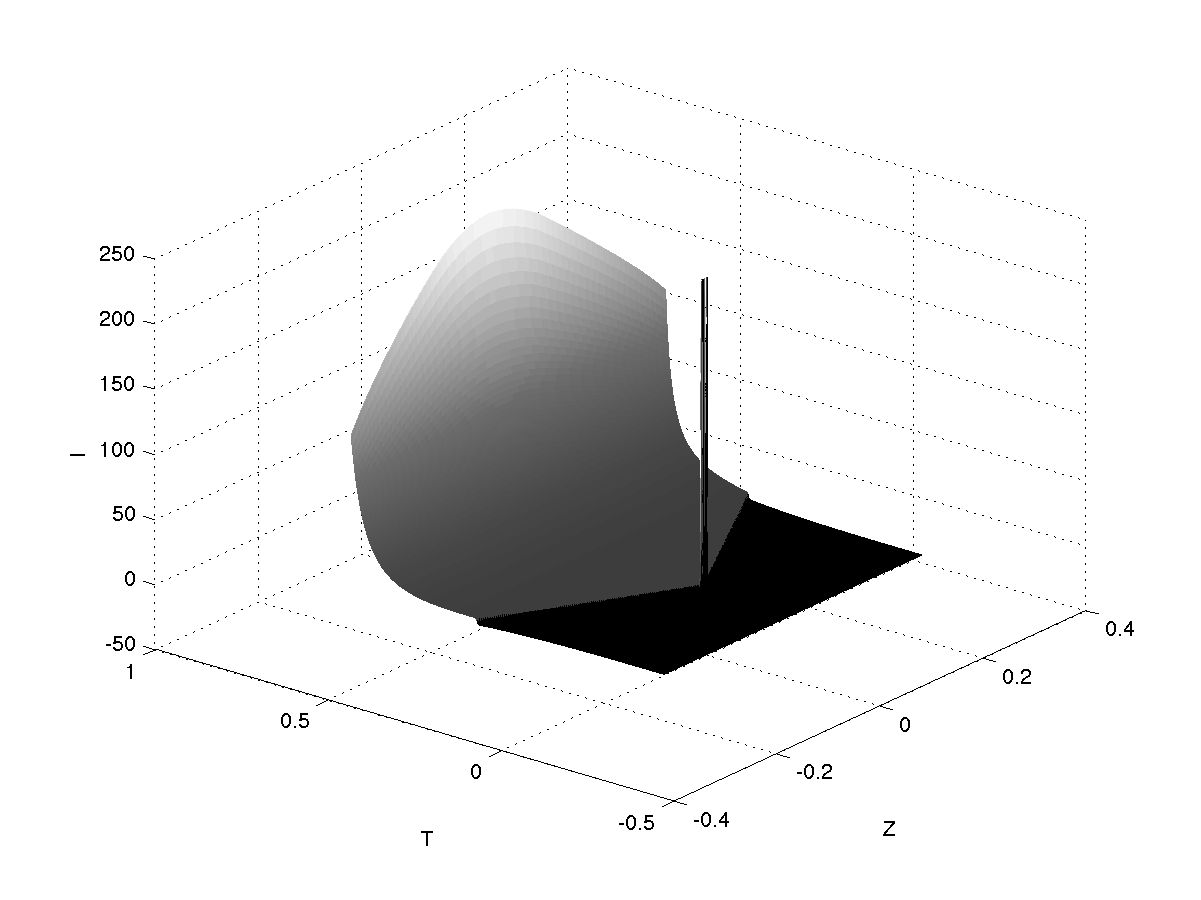}
  \caption{The curvature invariant $I$ for a run with amplitude $2^8$. The central peak which extends up to about $2^{16}$ has been capped.   \label{fig:KP_I_mesh}}
\end{figure}

Fig.~\ref{fig:KP_I_mesh} shows the curvature invariant $I$ for an ingoing wave amplitude of $2^8$, dependent on the coordinates $T$ and $Z$ determined from the characteristic coordinates $u$ and $v$, which have been computed along with the solution.
One can clearly distinguish three different regimes: the flat region consisting of the regions \R1, \R2, and \R3, in which no curvature is present, and region \R4, where a rapid divergence occurs across the full time-slice. These two regions are divided by an almost step-like increase along the $u=0$ and $v=0$ lines. Finally, we observe a central peak at $u=v=0$. This peak extends up to $65405.8\approx 2^{16}$, but has been capped in the figure to the same height as the maximum value at the final time.

In order to compare the computed space-time with the exact solution we need to relate the coordinate systems. The characteristic coordinates are already in agreement in the sense that the lines of constant $u$ and of constant $v$ agree with the lines $U=const$ and $V=const$. That is, they agree up to functions $U=f_1(u)$ and $V=f_2(v)$. To determine the functions $f_1$ and $f_2$, we use the discrete symmetry $z\mapsto -z$ or, equivalently, $u \leftrightarrow v$ of the system. Along the line $z=0$ in the $(u,v)$-plane we have $u=v$ in the numerical space-time and by construction the coordinate $t$ measures proper time $\tau=t$. Restricting to $U=V$ in the analytic solution, we find that
\[
\dd \tau^2 = \frac{2\left(1-2U^2\right)^{\tfrac32} }{1-U^2}\, \dd U^2,
\]
from which we can find $U(\tau)=V(\tau)$. Numerically, we compute $u(\tau)=v(\tau)$ along the central line with $z=0$, so that the function $f:=f_1=f_2$ is obtained implicitly by
$U(\tau) = f(u(\tau))$.

\begin{figure}[htb]
  \centering
  \includegraphics[width=0.8\textwidth]{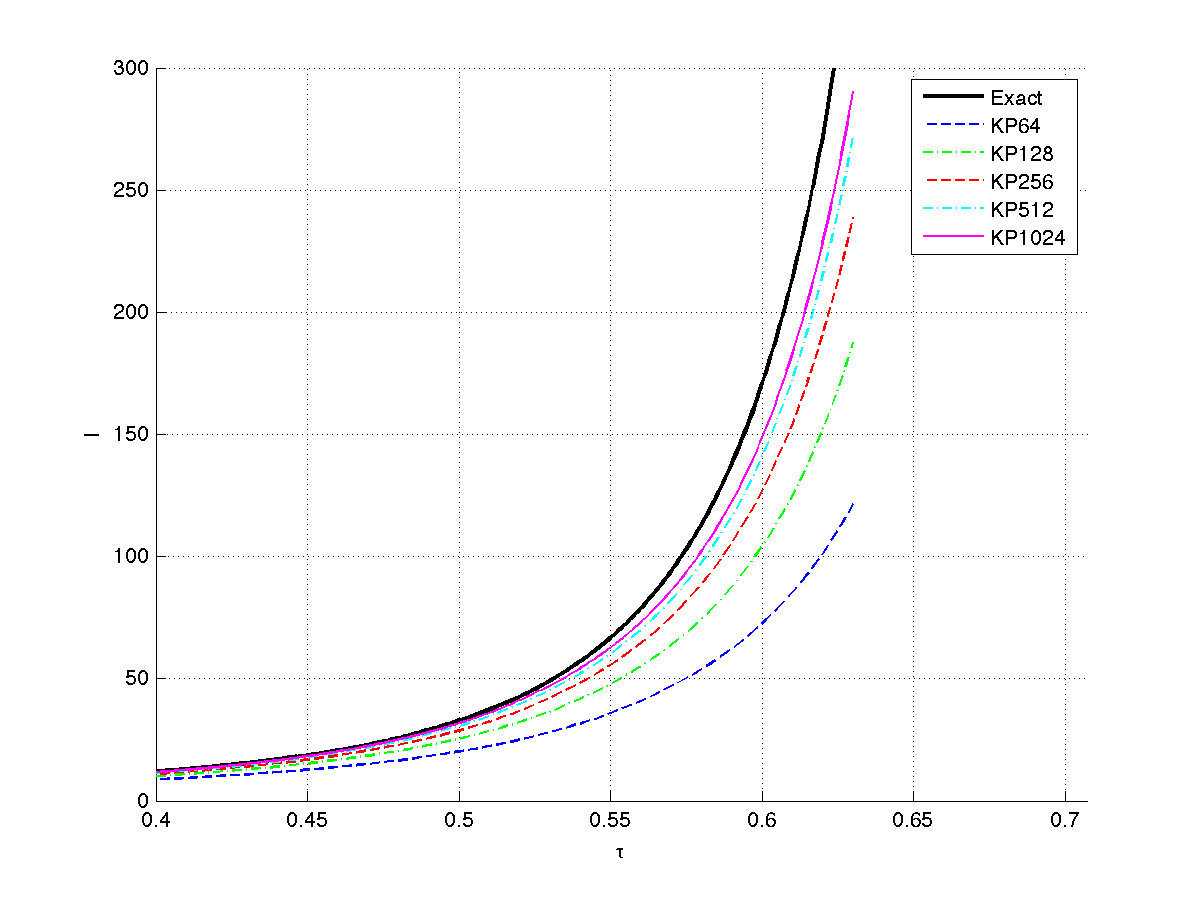}
  \caption{\label{fig:KP_I}
    The curvature invariant $I$ along the middle line as a function of proper time}
\end{figure}

In Fig.~\ref{fig:KP_I} we display the curvature invariant $I$ as a function of proper time $\tau$ along the $z=0$ line for the exact solution (thick, solid line), together with the approximations with increasing values of the amplitude $a$.
The value for the amplitude doubles between successive approximations. Table.~\ref{tab:relDelI} displays the relative distances $\tfrac{I(a)-I_{ex}}{I_{ex}}$ between the approximating and the exact solutions as a function of the amplitude $a$ at the final simulation time $t=0.88$.
\begin{table}[htb]
  \centering
  \begin{tabular}{c|rrrrr}
a &64 & 128 & 256 & 512 & 1024\\\hline
$\Delta I/I$& 0.6569    &0.4702    &0.3247    &0.2320    &0.1793 
\end{tabular}
\caption{The relative distance in the curvature invariant at $t=0.88$ as a function of the amplitude of the approximating solutions}
\label{tab:relDelI}
\end{table}

Evaluating the values in the table shows that the distance between successive approximations and the exact solution decreases roughly as $a^{-1/2}$, so the convergence to the exact impulsive solution is slow. Simultaneously, the computations become more and more challenging, since we need to resolve an increasingly steep function. Runs with amplitude of $10^4$ require a resolution of at least $10^6$ grid points.

\begin{figure}[htb]
  \centering
  \includegraphics[width=0.8\textwidth]{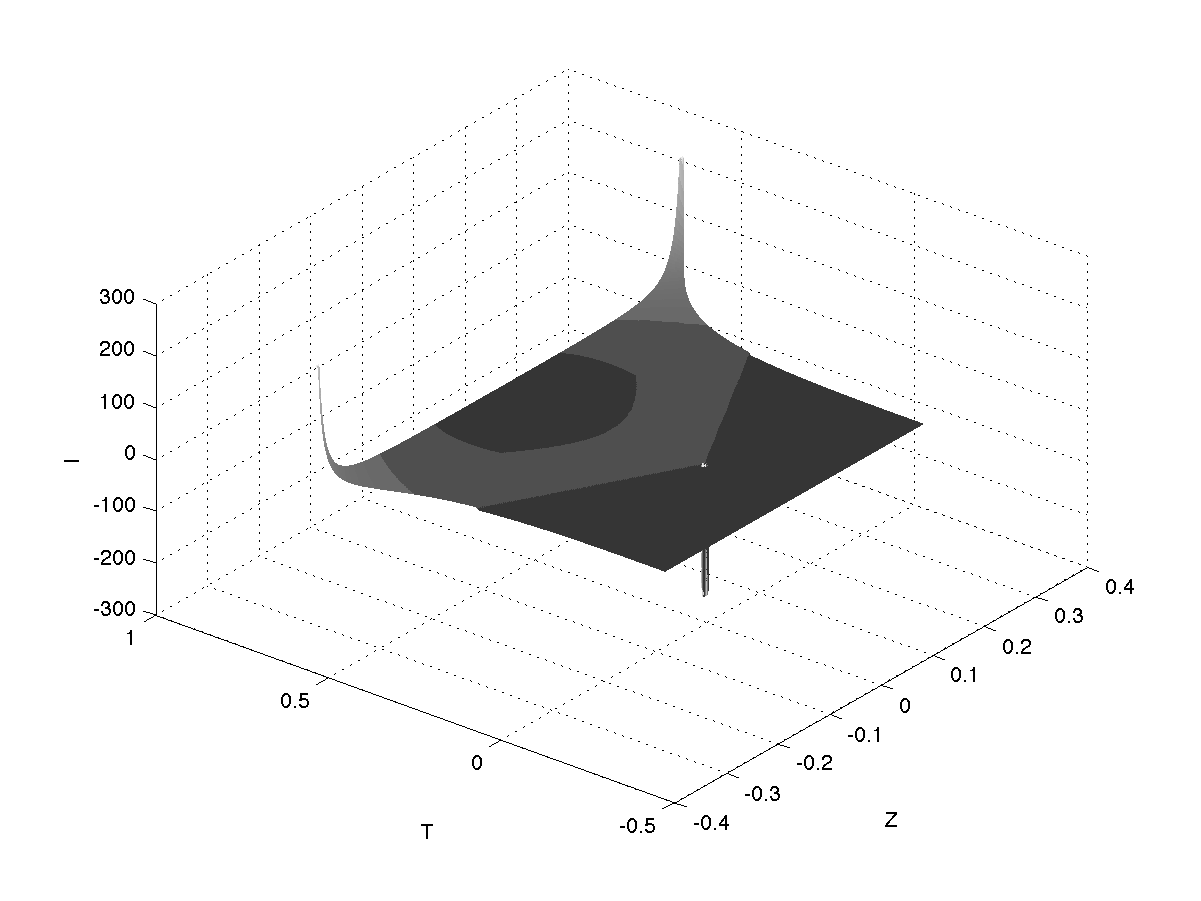}
  \caption{\label{fig:KP_I_pm} Curvature invariant for two colliding gravitational waves with opposite amplitudes}
\end{figure}

As another example, we show in Fig.~\ref{fig:KP_I_pm} the curvature invariant for the collision of two waves with opposite amplitudes, $\Psi(t,1)=-\Psi_4(t,-1)$.
In a linear theory, one would expect that the waves run through each other, interfering destructively, and then run off the grid, leaving nothing behind. In the non-linear case, we observe that the waves induce curvature, which stays around even after they have left the grid. In fact, the contribution to $I$ comes mostly from the component $\Psi_2$, which is determined entirely from the divergences and the shears, see~\eqref{eq:23}. In this case, however, the curvature scalar remains bounded inside the domain, but diverges along the boundaries where the waves were first noticeable. This results in an entirely different behaviour for the two opposite waves compared to the waves with equal amplitudes.

\section{Gravitational wave `ping-pong'}
\label{sec:grav-wave-ping}

For the second application of the code we set up the boundary conditions so that the system describes a gravitational wave `ping-pong'. Our main motivation for doing this is to test the implementation of the boundary conditions. We note that these boundary conditions are mathematically allowed in the sense that they give a well-posed IBVP. However, there is no physical motivation for them because there is no known physical process that could reflect gravitational waves. Nevertheless, it is interesting to see how the code can cope with these conditions. 

We feed an incoming wave into an otherwise flat region of space-time, through which it propagates until it hits the boundary on the other side of the computational domain. Here we can impose several different boundary conditions, which are compatible with the condition of maximal dissipation (see~\cite{Friedrich:1999dc})
\[
\Psi_{in}(t,\pm1) = \alpha(t) \, \bar\Psi_{out}(t,\pm1) + \psi_{in}(t), \qquad |\alpha(t)| \le 1.
\]
Here, $\Psi_{out}$ is the outgoing wave, which is partly transmitted and partly reflected into an ingoing wave profile according to the complex time-dependent reflection coefficient $\alpha(t)$, which describes the (time-dependent) reflection properties. The function $\psi_{in}(t)$ is a profile for a wave coming in additionally from the left. 

The incoming left-moving wave enters through the right boundary with the profile
\[
\psi_0(t) =
\begin{cases}
a \sin(b t)^8 &\text{for } 0 \le t \le \frac\pi{b}\\
  0 & \text{otherwise} 
\end{cases},
\]
for appropriate values of the constants $a$ and $b$. The wave propagates through the computational domain and is reflected back into the domain at the left boundary with $\alpha(t)=\i$ and $\psi_4(t)=0$. When it hits the right boundary again, we reflect it back in with the same phase shift, so that the boundary condition on the right end is
\[
\Psi_0(t,1) = -\i\bar\Psi_4(t,1) + \psi_0(t), \qquad \text{for } t \ge 0.
\]
Thus, every reflection changes the polarisation of the wave by $\frac\pi2$.

\begin{figure}[htb]
  \centering
  \includegraphics[width=\textwidth]{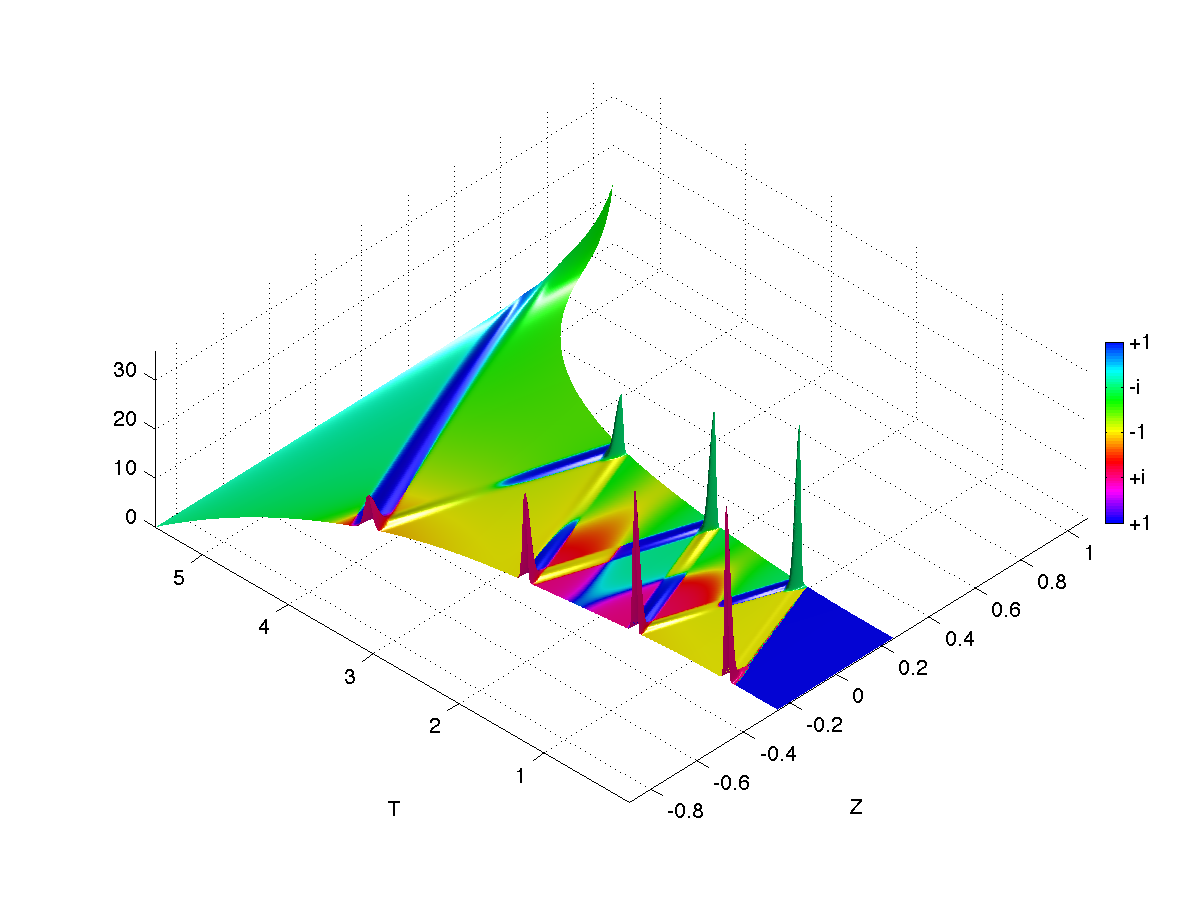}
  \caption{The complex curvature invariant $I$ for gravitational wave `ping-pong'. The height of the surface is $|I|$, while the colour coding corresponds to the complex phase of the function~$I$. This particular plot was obtained using $a=6$ and $b=20$.}
  \label{fig:I}
\end{figure}

We will discuss these `ping-pong' systems in another paper in more detail. Here, we want to restrict ourselves to some more or less obvious observations. We refer to~Fig.~\ref{fig:I}, where we plot a representation of the curvature scalar $I= \Psi_0\Psi_4 + 3\Psi_2^2$. The height of the surface shows the absolute value $|I|$, while the coloring corresponds to the complex phase of $I$. 
\begin{itemize}
\item Depending on the parameters $a$ and $b$ of the ingoing wave profile, the wave bounces back and forth several times until finally a divergence occurs. This is due to the divergence in $\Psi_0$, and it  describes a genuine curvature singularity and not a coordinate phenomenon, since both of the Weyl invariants $I$ and~$J$ (see~\cite{Penrose:1984wm}) diverge. By decreasing the amplitude of the ingoing wave, we can increase the number of bounces, but in every run so far there has always been a divergence on the boundary, through which the initial wave traveled into the computational domain. The curvature invariant vanishes until the ingoing wave hits the opposite boundary for the first time. The reason for this is simply that $\Psi_4$, $\rho'$, and $\sigma'$ vanish initially so that $I$, which is made up from products $\rho\rho'$, $\sigma\sigma'$, and $\Psi_0\Psi_4$ also vanishes. Only when the wave is reflected for the first time is there a $\Psi_4$ component going to the right, which generates $\rho'$ and $\sigma'$, and therefore produces a non-vanishing $I$.
\item During the reflection phase, when the ingoing and outgoing waves have a non-vanishing overlap, we see `spikes' in the curvature invariants near the boundaries, indicating a period of intense interaction. The spikes are due to the contributions from the $\Psi_0$ and $\Psi_4$ profiles directly, while the lower values of $I$ are caused by the indirect interaction between the waves via the spin coefficients. The wave amplitudes, and therefore the spikes, decrease in amplitude over time. Why this happens is unclear.
\item During the first bounces, the phases of $\Psi_0$ and $\Psi_4$ go through the sequence $0\to\frac\pi2 \to \pi \to \frac{3\pi}2 \to 2\pi \to \cdots$ as expected, but the more bounces there are the more obvious it becomes that there is a secular, non-periodic phase shift, which must be caused by the interaction between the wave and the increasingly curved `background', on which it propagates.
\item During the evolution, the geometry of the space-time region changes. We demonstrate this by computing the proper length along the intervals given by $t=const$. These increase over time. The travel times from one end of the interval to the other end increase accordingly.
\end{itemize}

\section{Summary}
\label{sec:summary}

In this paper, we have presented a numerical implementation of the Friedrich-Nagy gauge conditions for the Einstein equations in the symmetry-reduced context of plane waves. We have shown that one can use the resulting algorithm to solve an IBVP for the Einstein equations in a numerically stable way. We have used the code to reproduce the well-known Khan-Penrose solution, which describes colliding impulsive waves. Another application of the code is the simulation of a gravitational wave that is reflected with a phase shift back and forth between two boundaries. We observe that this setup leads to a curvature singularity. Apparently, this behaviour is independent of the boundary conditions used. 

The divergence seems to be driven entirely by the behaviour of $\rho$ or $\rho'$ along the boundaries. The equations for $\rho$ and $\sigma$ on the right boundary are essentially Sachs' optical equations, and their solutions are driven by curvature component $\Psi_0$. As soon as this becomes non-zero the boundary gets a non-vanishing $\rho$ and $\sigma$. The divergence of $\rho$ is therefore to be expected. The plane symmetry of the space-time leads to the effect that $\Psi_2$ is determined only in terms of the shears and divergences (see~(\ref{eq:23})), which is registered in the curvature invariant. So, unless the evolution is such that the combinations in $\Psi_2$ cancel each other, the divergence in $\rho$ along the boundary will inevitably lead to a curvature singularity.

These observations seem to be in line with a theorem proved by Tipler~\cite{Tipler:1980fa}, which states that a space-time with plane symmetry is null incomplete if there is at least one point $p$, at which one of the Newman-Penrose quantities, $(\Psi_0,\Phi_{00},\sigma)$ (or their primed versions) is non-zero, provided that the null-convergence condition holds, and that $p$ is contained in an (invariant) Cauchy surface which is non-compact in the direction perpendicular to the symmetry generators. 

It would be interesting to see whether this situation has any similarity to the recently discovered weakly turbulent instability~\cite{Bizon:2011kz} of anti-de Sitter space, where a massless scalar field coupled to the Einstein equations propagates on an asymptotically anti-de Sitter space-time. Successive reflections at time-like infinity transfer energy from low to high frequency modes, which ultimately collapse. To answer this question, one needs to understand the reflecting plane wave systems better, and more detailed numerical studies must be carried out.

\appendix

\section{Exact expressions for the Khan-Penrose solution}
\label{sec:exact-kp}

In region \R4 we have $R=\sqrt{1-U^2}$, $W=\sqrt{1-V^2}$ and $T=\sqrt{1-U^2-V^2}$. We also define $M:=(U V + R W)$ and $N^2:=U V R W$. With these functions the Weyl curvature components and the curvature invariants $I$ and $J$ are
\begin{equation}
  \label{eq:22}
  \begin{gathered}
  \Psi_0 = \frac1R \delta(V) + 3U \frac{R^2}{W} \frac{M^3}{T^7},\quad
  \Psi_2 =  \frac{M^2}{T^7} \left( M^2 - N^2 \right),\quad
  \Psi_4 = \frac1W\delta(U) + 3V \frac{W^2}{R} \frac{M^3}{T^7},\\
  I = \delta(U) \delta(V) + 3 (M^4 + M^2N^2  + N^4 )\frac{M^4}{T^{14}},\\
  J = \delta(U) \delta(V) - (M^2 - N^2)(M^4 -11 M^2N^2  + N^4 )\frac{M^6}{T^{21}}.
  \end{gathered}
\end{equation}
In obtaining these expressions it is understood that terms of the form $x\delta(x)$ and similar are put equal to zero.

\acknowledgements
The authors are grateful to H. Friedrich for discussions. This work was supported by the Marsden Fund Council from Government funding, administered by the Royal Society of New Zealand. The authors wish to acknowledge the contribution of NeSI high-performance computing facilities to the results of this research. This material is based upon work supported by the National Science Foundation under Grant No. 0932078 000, while JF was in residence at the Mathematical Sciences Research Institute in Berkeley, California, during the winter semester of 2013. \jf{Please notify NeSI (\url{pubs@nesi.org.nz}) when you make an acknowledgment, which assists us greatly with record keeping.}


\begin{thebibliography}{10}
\providecommand{\url}[1]{{#1}}
\providecommand{\urlprefix}{URL }
\expandafter\ifx\csname urlstyle\endcsname\relax
  \providecommand{\doi}[1]{DOI~\discretionary{}{}{}#1}\else
  \providecommand{\doi}{DOI~\discretionary{}{}{}\begingroup
  \urlstyle{rm}\Url}\fi

\bibitem{Bartnik:1997jh}
Bartnik, R.: Einstein equations in the null quasispherical gauge.
\newblock Class. Quantum Grav. \textbf{14}(8), 2185--2194 (1997)

\bibitem{Bizon:2011kz}
Bizo{\'n}, P., Rostworowski, A.: On weakly turbulent instability of anti-de
  {S}itter space.
\newblock Phys. Rev. Lett. \textbf{107} (2011)

\bibitem{Carpenter:1993uu}
Carpenter, M.H., Gottlieb, D., Abarbanel, S.: The stability of numerical
  boundary treatments for compact high-order finite-difference schemes.
\newblock J. Comp. Phys. \textbf{108}(2), 272--295 (1993)

\bibitem{Carpenter:1993wh}
Carpenter, M.H., Gottlieb, D., Abarbanel, S.: Stable and accurate boundary
  treatments for compact, high-order finite-difference schemes.
\newblock Appl. Numer. Math. \textbf{12}(1-3), 55--87 (1993)

\bibitem{Frauendiener:2004te}
Frauendiener, J.: Conformal infinity.
\newblock Living Rev. Relativity \textbf{7}, 2004--1, 82 pp. (electronic)
  (2004)

\bibitem{Friedrich:1995uf}
Friedrich, H.: Einstein equations and conformal structure: {E}xistence of
  anti-de {S}itter-type space-times.
\newblock J. Geom. Phys. \textbf{17}, 125--184 (1995)

\bibitem{Friedrich:1999dc}
Friedrich, H., Nagy, G.: The initial boundary value problem for {E}instein's
  vacuum field equations.
\newblock Commun. Math. Phys. \textbf{201}(3), 619--655 (1999)

\bibitem{Griffiths:1991vx}
Griffiths, J.B.: Colliding plane waves in general relativity.
\newblock Oxford University Press, USA (1991)

\bibitem{Khan:1971ew}
Khan, K.A., Penrose, R.: Scattering of two impulsive gravitational plane waves.
\newblock Nature \textbf{229}(5281), 185--186 (1971)

\bibitem{Kijowski:1997wt}
Kijowski, J.: A simple derivation of canonical structure and quasi-local
  hamiltonians in general relativity.
\newblock Gen. Rel. Grav.  (1997)

\bibitem{Lehner:2004ha}
Lehner, L., Neilsen, D., Reula, O., Tiglio, M.: The discrete energy method in
  numerical relativity: towards long-term stability.
\newblock Class. Quantum Grav. \textbf{21}(24), 5819--5848 (2004)

\bibitem{Matzner:1984bx}
Matzner, R.A., Tipler, F.J.: Metaphysics of colliding self-gravitating plane
  waves.
\newblock Phys. Rev. D \textbf{29}(8), 1575--1583 (1984)

\bibitem{Newman:1962ue}
Newman, E.T., Penrose, R.: An approach to gravitational radiation by a method
  of spin coefficients.
\newblock J Math Phys \textbf{3}(3), 566--578 (1962)

\bibitem{Penrose:1984wm}
Penrose, R., Rindler, W.: Spinors and {S}pacetime: {T}wo-spinor calculus and
  relativistic fields, vol.~1.
\newblock Cambridge University Press, Cambridge (1984)

\bibitem{Sarbach:2011tq}
Sarbach, O., Tiglio, M.: Continuum and discrete initial-boundary-value problems
  and {E}instein's field equations.
\newblock Living Rev. Relativity pp. 1--182 (2011)

\bibitem{Schnetter:2006ku}
Schnetter, E., Diener, P., Dorband, E.N., Tiglio, M.: A multi-block
  infrastructure for three-dimensional time-dependent numerical relativity.
\newblock Class. Quantum Grav. \textbf{23}(16), S553--S578 (2006)

\bibitem{Strand:1994ef}
Strand, B.: Summation by parts for finite difference approximations for d/dx.
\newblock J. Comp. Phys. \textbf{110}(1), 47--67 (1994)

\bibitem{Tamburino:1966vd}
Tamburino, L.A., Winicour, J.H.: Gravitational fields in finite and conformal
  {B}ondi frames.
\newblock Phys. Rev.  (1966)

\bibitem{Tipler:1980fa}
Tipler, F.J.: Singularities from colliding plane gravitational waves.
\newblock Phys. Rev. D \textbf{22}(12), 2929--2932 (1980)

\bibitem{Winicour:2001tn}
Winicour, J.: Characteristic evolution and matching.
\newblock Living Rev. Relativity \textbf{4}, 2001--3-- 65 pp. (electronic)
  (2001)

\end{thebibliography}
\end{document}